\documentstyle[prd,aps,floats]{revtex}

\begin{document}
\preprint{astro-ph/0004296}
\draft

\input epsf

\renewcommand{\topfraction}{0.99}
\renewcommand{\bottomfraction}{0.99}

\twocolumn[\hsize\textwidth\columnwidth\hsize\csname
@twocolumnfalse\endcsname

\title{Black hole constraints on the running-mass inflation model}
\author{Samuel M.~Leach$^{1,2}$, Ian J.~Grivell$^1$ and Andrew 
R.~Liddle$^{1,2}$}
\address{$^1$Astrophysics Group, The Blackett Laboratory, Imperial College, 
London SW7 2BZ, United Kingdom\\
$^2$Astronomy Centre, University of Sussex, Brighton BN1 9QJ, United Kingdom 
(present address)} 
\date{\today} 
\maketitle
\begin{abstract}
The running-mass inflation model, which has strong motivation from particle 
physics, predicts density perturbations whose spectral index is strongly 
scale-dependent. For a large part of parameter space the spectrum rises sharply 
to short scales. In this paper we compute the production of primordial black 
holes, using both analytic and numerical calculation of the density perturbation 
spectra. Observational constraints from black hole production are shown to 
exclude a large region of otherwise permissible parameter space.
\end{abstract}

\pacs{PACS numbers: 98.80.Cq \hfill astro-ph/0004296}

\vskip2pc]

\section{Introduction}

Particle physics models of inflation based on supergravity theories are plagued
by the so-called $\eta$-problem \cite{LR}, which states that the mass-squared of 
any
scalar field, including the putative inflaton field, is typically of order $H^2$ 
($H$
being the Hubble parameter) which ruins slow-roll inflation.  An elegant
proposal to circumvent this is the running-mass model of inflation,
introduced by Stewart \cite{S1,S2}, where the flatness of the potential arises
because of the quantum corrections, which serve to flatten the potential over a
significant region where inflation can then take place.

Because the flatness is brought about by a cancellation of the intrinsic 
curvature of the potential against the quantum corrections, only a limited 
portion of the potential can support slow-roll inflation, as is necessary to 
generate the approximately flat power spectrum seen by the COBE satellite. Well 
away from COBE scales, one typically expects to see dramatic deviations from 
near scale-invariance as the slow-roll regime breaks down. Copeland et 
al.~\cite{CGL} examined the possibility that this breakdown of scale-invariance 
might be detectable through scale-dependence of the spectral index of primordial 
perturbations, and more recently the model has been extensively explored by 
Covi, Lyth and collaborators \cite{CLR,CL,LC} in a series
of papers investigating its  viability both from a theoretical
standpoint and in confrontation against large-scale structure data.

In this paper we examine constraints arising from the more radical departures 
from scale-invariance which may take place towards the end of inflation. In much 
of parameter space, the spectrum rises sharply on short scales, which can give 
rise to production of primordial black holes (PBHs). These are strongly 
constrained by 
observation and, as we will see, a significant region of otherwise viable 
parameter space is excluded.

\section{The running-mass model}

Whether or not a potential $V(\phi)$ can support slow-roll inflation can be 
judged via the slow-roll parameters \cite{LL}
\begin{equation}
\label{eqn:epsilon_defn}
\label{eqn:eta_defn}
\epsilon_{{\scriptscriptstyle V}} \equiv \frac{1}{2}M^2_{{\rm 
P}}\left(\frac{V'}{V}\right)^2 \quad ; \quad
\eta_{{\scriptscriptstyle V}} \equiv M^2_{{\rm P}}\frac{V''}{V} \,, 
\end{equation}
where primes are $\phi$ derivatives. When the slow-roll parameters are much less 
than unity, slow-roll inflation can proceed and gives rise to perturbations with 
an approximately scale-invariant spectrum.

Within the context of softly-broken global supersymmetry, the false
vacuum dominated potential 
\begin{equation}
\label{eqn:V1}
V= V_0\left[1 - \frac{1}{2} \, \mu^2 \, \frac{\phi^2}{M_{{\rm P}}^2} + ...
	\right] \,, 
\end{equation}
arises naturally \cite{LR}. However it will not in general lead to slow-roll 
inflation, because supergravity corrections lead to $|\mu^2| = 
|\eta_{{\scriptscriptstyle V}}| \simeq 1$ in Planck units. In the scenario 
proposed by Stewart~\cite{S1,S2}, the inflaton has gauge couplings to vector or 
chiral superfields, and one-loop quantum corrections flatten the
potential, corresponding to a running of the effective mass with the scalar 
field value
\begin{equation}
\mu^2 \equiv \mu^2\left[\mu^2_0,A_0,\tilde{\alpha}_0 \ln \frac{\phi}{M_{{\rm 
P}}}\right] 
\,,
\end{equation}
where $\mu^2_0$ represents the inflaton mass squared, $A_0$ is the mass 
squared of the gaugino appearing in the loop, both evaluated at the
Planck scale, and $\tilde{\alpha}_0$ is a gauge coupling times a group theoretic 
factor
which may be positive (in the case of asymptotic freedom) or negative (in the 
opposite case).  The functional form 
of $\mu^2(\phi)$ is obtained by solving the relevant renormalization group
equations. For definiteness, we consider the inflaton potential 
\cite{S1,S2,CLR,CL}
\begin{equation}
\frac{V}{V_0}=1-\frac{\phi^2}{2 M^{2}_{{\rm
	P}}}\left\{\mu_0^2+A_0\left[1-\frac{1}{\left(1+
	\tilde{\alpha}_0 \ln \frac{\phi}{M_{{\rm P}}}\right)^2}
	\right]\right\}.\label{eqn:inf_potential}
\end{equation}
The term proportional to $A_0$ vanishes when $\phi=M_{{\rm P}}$ and 
the potential reverts to the form of Eq.~(\ref{eqn:V1}). Far below the 
Planck scale, the desired cancellation occurs between the $\mu^2_0$ and
$A_0$ pieces allowing slow-roll inflation to occur.

An attractive feature of this model is that its parameters take on
natural values; for $\tilde{\alpha}_0>0$
we have $A_0$ and $\mu_0^2$ both positive and of order unity, while for  
$\tilde{\alpha}_0<0$ we have $A_0$ negative and the model is subject to the
constraint $\left|A_0\right| > \mu_0^2 + 1$, which is applied to
ensure that the inflaton mass changes sign before reaching the end of 
inflation. Full details of the model can be found in Refs.~\cite{CLR,CL}.

\begin{figure}[t]
\centering 
\leavevmode\epsfysize=6cm \epsfbox{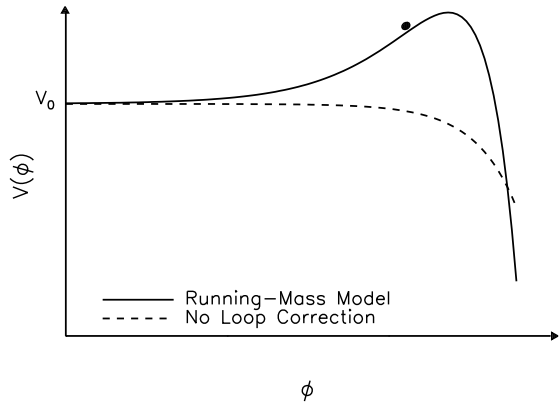}\\
\caption[fig1]{\label{fig:pot_plot} A sketch of the potential. The inflaton
starts near the maximum with $\eta_{{\scriptscriptstyle V}}$ negative. As it 
rolls towards the origin
the mass passes through zero and $\eta_{{\scriptscriptstyle V}}$ grows to 1. For 
the purposes of this diagram the bump is greatly exaggerated; in reality the 
potential is extremely flat.}
\end{figure}

The inflaton field starts off near the maximum of the potential\footnote{Note 
that initial conditions near the maximum of the potential are well motivated by 
the `topological inflation' idea~\cite{V}; the initial conditions inevitably 
have 
the field on different sides of the maximum in different regions of space, and 
hence crossing the maximum in the interpolating regions.} and rolls towards the
origin, corresponding to model~$(i)$ of Ref.~\cite{CL} and as shown in
Fig.~\ref{fig:pot_plot}. In the case of $\tilde{\alpha}_0>0$, the potential has 
an
unphysical pole at $\ln(\phi/M_{{\rm P}})=1/\tilde{\alpha}_0$ and should not be 
trusted
in this region of strong coupling.  It is not necessary, though, to evolve 
$\phi$
to such small field values.

Throughout this calculation we use the Hubble-slow-roll parameters,
defined as \cite{LPB}
\begin{eqnarray}
\label{eqn:hubble_defn}
&&\epsilon \equiv 2 M^2_{{\rm P}}\left(\frac{H'}{H}\right)^2 \quad ; \quad
\eta \equiv 2 M^2_{{\rm P}}\frac{H''}{H} \,; \nonumber \\
&&\xi \equiv  2 M^2_{{\rm P}}\left(\frac{H'H'''}{H^2}\right)^{1/2}, 
\end{eqnarray}
where the fundamental quantity is now taken to be the Hubble parameter $H$ and 
its derivatives, rather than the potential.

The value of $\eta$ starts off negative near the maximum of the potential
and runs through zero until the end of slow-roll inflation is reached, which we 
define to be
\begin{equation}
\eta(\phi_{{\rm end}})=1.\label{eqn:eoi_defn}
\end{equation}
At this point the inflaton potential is still dominated by the $V_0$ term, so 
it is assumed that the field must decay via some hybrid
inflation mechanism when $\phi$ falls below a critical value $\phi_{{\rm c}}$, 
in order that reheating occurs to restore the standard cosmology.

\section{Computing the perturbation spectra}

Our main focus is the perturbations near the end of inflation, 
where the slow-roll approximation will be poor. It is therefore imperative that 
the accuracy of calculations is checked numerically. There are two aspects to 
this; numerical calculation of the classical scalar field dynamics, and 
numerical calculation of the perturbation equations. 

The numerical calculation of the classical evolution is important in determining 
which part of the potential generates the perturbations seen on cosmological 
scales. When the slow-roll approximation begins to break down, commonly-used 
expressions such as that for the number of $e$-foldings $N$ can lose their 
accuracy and the numerical evolution can lead to some corrections to the 
analytic results.
In general, $N$ and the wavemode $k$ leaving the horizon at that epoch are 
related by
\begin{equation}
N\left(k\right)=N_{{\rm COBE}}- \ln\frac{k}{k_{{\rm COBE}}} \,,
 \label{eqn:Nk_relation}
\end{equation}
where $N_{{\rm COBE}}$ is defined throughout as the number of $e$-foldings
before the end of inflation when our present Hubble radius, in
comoving units, equalled the Hubble radius during inflation. This expression
neglects the variation of $H$, which is valid as long as 
$\epsilon_{{\scriptscriptstyle V}}$ is small even when 
$\eta_{{\scriptscriptstyle V}}$ is not.

To numerically compute the perturbations, we use the Mukhanov formalism
\cite{Muk,MFB} as described by Grivell and Liddle \cite{GL1}, to which we refer
the reader for details.  The approach involves a numerical solution both of the
classical homogeneous equations of motion and of the equations describing linear
perturbations, with the full power spectrum being built up mode-by-mode.  In
order to evaluate the power spectrum, one has to follow the modes until they are
well outside the horizon, where their amplitude becomes constant.  This becomes
problematic once one reaches very close to the end of inflation, where this
asymptotic regime is not reached.  In fact, in the case of the running-mass
model it proves difficult to numerically evolve the mode evolution much more
than around one $e$-folding beyond the $\eta = 1$ point, since the slow-roll
parameters $\eta$ and $\xi^2$ are growing so rapidly at this point.

\begin{figure}[t]
\centering 
\leavevmode\epsfysize=6cm \epsfbox{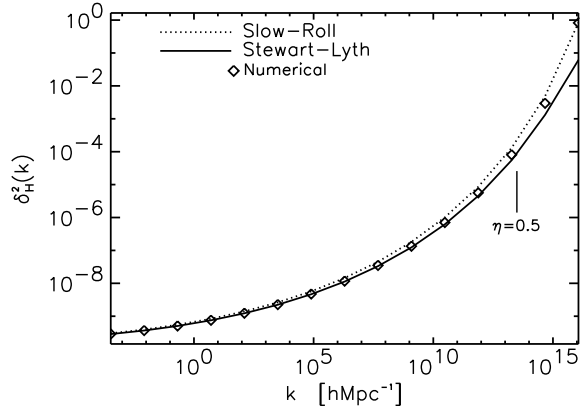}\\
\caption[fig1]{\label{fig:ex_spec} An example power spectrum, taking 
$\tilde{\alpha}_0 = 
0.01$, $A_0 = 1.0$, $\mu_0^2 = 0.5$ and $N_{{\rm COBE}} = 45$. The 
numerical calculation of the perturbation amplitude breaks down towards the end 
of inflation. The end point of the power spectrum ($k_{{\rm end}} = e^{N_{{\rm 
COBE}}} \, h/3000 \, {\rm Mpc}^{-1}$) is defined to be where $\eta = 1$. The 
position where $\eta = 0.5$ is shown for comparison.}
\end{figure}

However, as long as one uses the numerical solution for the 
classical background evolution, it turns out we do not need to compute the 
perturbations numerically; the extended slow-roll approximation of Stewart and 
Lyth \cite{SL} proves perfectly adequate. This gives the perturbation amplitude 
as
\begin{equation}
\delta_{{\rm H}}(k) \simeq \left[1-(2C+1)\epsilon+C\eta\right]
 \frac{1}{10\pi M_{{\rm P}}^2}\left.\frac{H^2}{\left|H'\right|}\right|_{k=aH},
 \label{eqn:sl_approx} 
\end{equation}
where $C = -2 + \ln 2 + b \simeq -0.73$ and \emph{b} is the
Euler--Mascheroni constant, and gives sufficiently accurate results even when 
$\eta$ becomes large. Fig.~\ref{fig:ex_spec} shows a comparison of numerical 
simulation with the slow-roll and Stewart--Lyth predictions.

The error in the Stewart--Lyth expression is expected
to be ${\mathcal O}(\xi^2)$.  It is known
to underestimate the perturbation amplitude when $\eta \simeq 1$ and
$\epsilon \ll 1$ \cite{GL1}, which makes our PBH constraints
conservative. A further consequence of the smallness of $\epsilon$,
\begin{equation}
\epsilon_{{\scriptscriptstyle  V}} =
    \frac{1}{2} \, \eta_{{\scriptscriptstyle V}}^2 \, \frac{\phi^2}{M_{{\rm 
P}}^2} 
\,, \label{eqn:epsilon_suppression}
\end{equation}
is that the gravitational waves from this model are strongly
suppressed \cite{LL}.

We use the COBE normalization scheme of Ref.~\cite{BLW}, setting $\Omega_0 =
0.35$ and $\Omega_{\Lambda}= 0.65$.  In fact the normalization of the potential
is very nearly independent of the cosmological parameters, since the temperature
anisotropies are, with the exception of the integrated Sachs--Wolfe effect, laid
down at the redshift of decoupling, long before the cosmological constant is
important.  In this scheme, normalization occurs at the scale $k = 7a_0H_0 =
7h/3000 \, {\rm Mpc}^{-1}$.

A typical power spectrum is shown in Fig.~\ref{fig:ex_spec}, taken from a region
of parameter space that we will show to be excluded by PBH constraints. The 
scale-dependence of the spectral index is clearly seen. By
virtue of the $N(k)$ relation of Eq.~(\ref{eqn:Nk_relation}), a change in
$N_{{\rm COBE}}$ corresponds simply to a \emph{translation} of the power
spectrum to a new end point given by $k_{{\rm end}}=e^{N_{{\rm COBE}}}\,h/3000
\, {\rm Mpc}^{-1}$ (though it must be renormalized to COBE) .  A change in the
condition for the end of slow-roll inflation also results in a translation and
renormalization of the power spectrum.  As we shall see in
Section~\ref{sec:pbh}, the condition for the end of slow-roll inflation,
Eq.~(\ref{eqn:eoi_defn}), is one of two parameters that affect the severity of
the PBH constraints the most, the other being $N_{{\rm COBE}}$.

In Section~\ref{sec:pbh} we consider the cases $N_{{\rm COBE}} = 45$, 
which corresponds to instant reheating after the end of slow-roll
inflation, and $N_{{\rm COBE}} = 25$, which may result if the end of 
slow-roll is followed by a bout of fast-roll and/or thermal inflation.

\begin{figure*}[t]
\centering 
\leavevmode \epsfysize=6.0cm \epsfbox{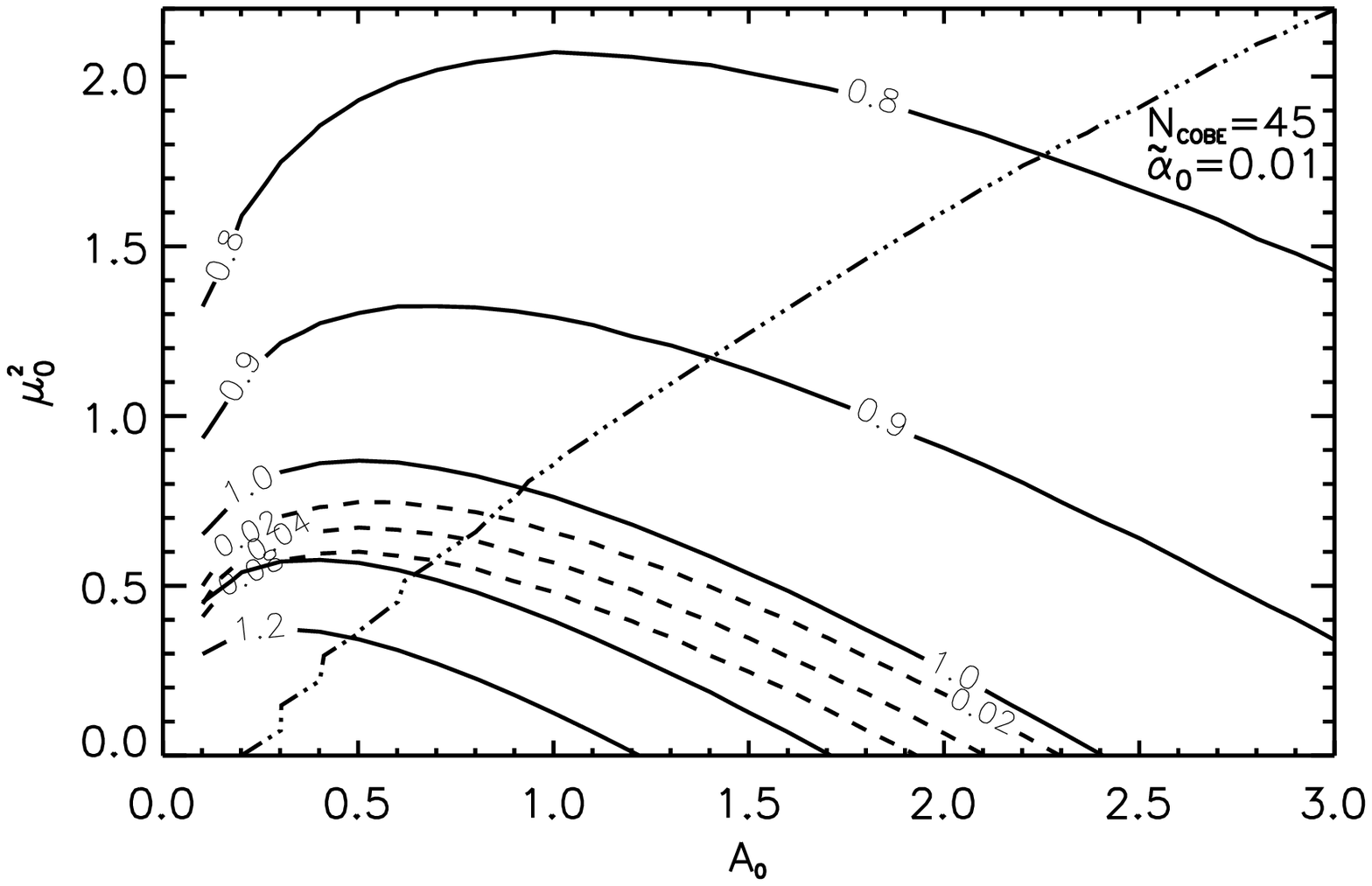}
\epsfysize=6.0cm \epsfbox{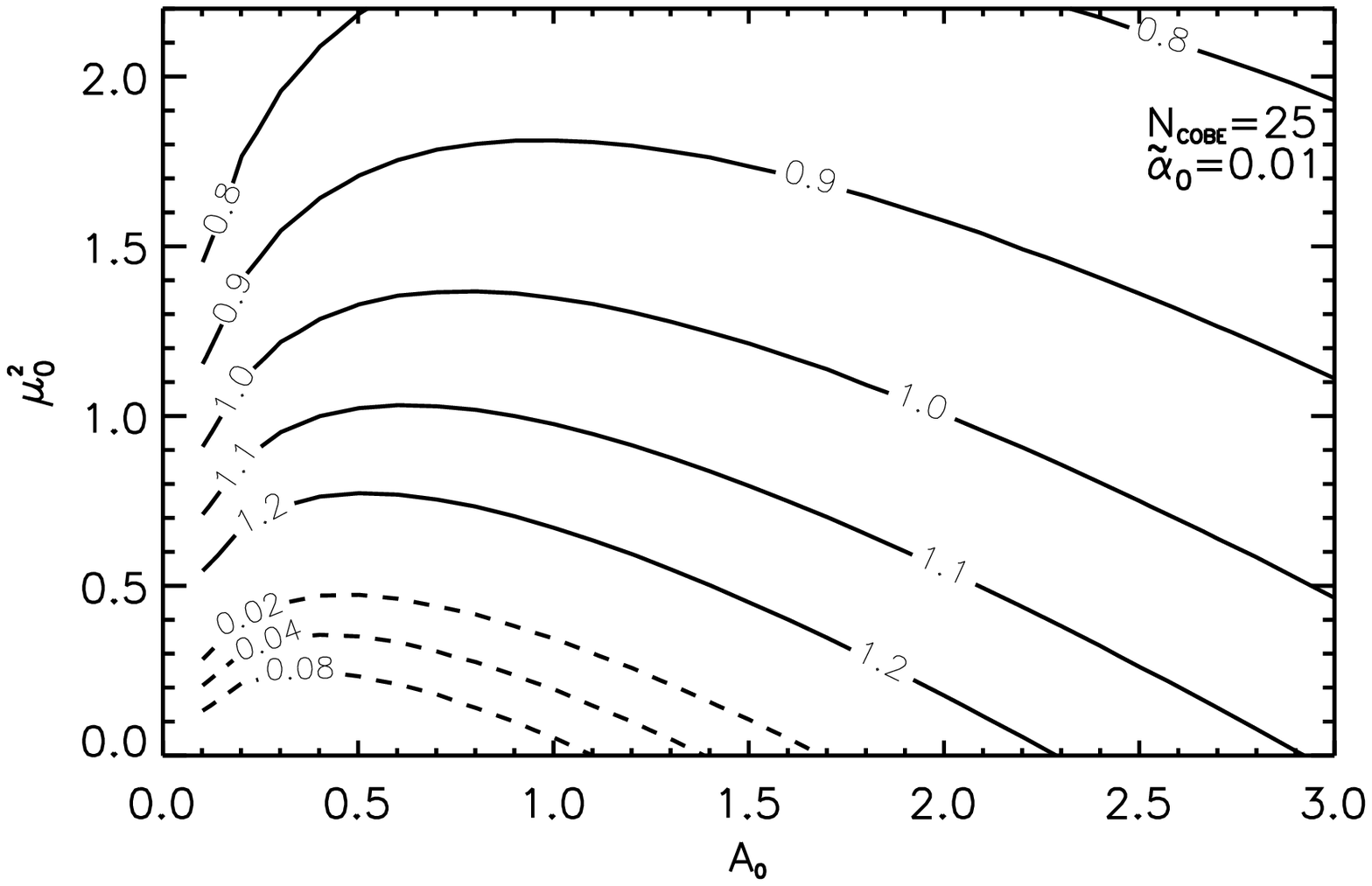} 
\caption[fig1]{\label{fig:alpha_pos} Parameter space constraints for 
$\tilde{\alpha}_0 = 
0.01$ for two choices of $N_{{\rm COBE}}$. The solid lines show the spectral 
index on COBE scales. The dashed lines are 
$\sigma_{{\rm hor}} = 0.02, 0.04, 0.08$ contours,
where $\sigma_{{\rm hor}}$ has been evaluated at the end of slow-roll inflation. 
Parameter space below this region is excluded, and these models will
violate the bound on $\sigma_{{\rm hor}}$ \emph{before} the end of
slow-roll inflation. From Eq.~(\ref{eqn:instant_r}), instant reheating requires
$V_0/M^4_{{\rm P}} \gtrsim 10^{-36}$ for $N_{{\rm COBE}}=45$, indicated by
the dot-dashed contour; the parameter space above this contour is excluded.}
\end{figure*} 

\section{Constraints from primordial black hole production}
\label{sec:pbh}

\subsection{The black hole constraint}

The astrophysical details of PBH constraints have been studied in detail 
elsewhere \cite{GL2}. Over a wide range of mass scales, the observational 
constraint on the black hole formation rate is that no more than around 
$10^{-20}$ of the mass of the Universe can be channeled into black holes.
For our purposes it is sufficient to ignore the details of the constraints, and 
simply adopt this level, as the black hole production rate is enormously 
sensitive to the amplitude of perturbations.

In computing the black hole formation rate, the quantity which is of interest is 
the matter dispersion $\sigma$, which is defined, in the usual way (see 
e.g.~Ref.~\cite{LL}), as the 
matter
distribution smoothed over some length scale $R$,
\begin{equation}
\sigma^2(R,t) = \left(\frac{10}{9}\right)^2 \int_{0}^{\infty} 
\left(\frac{k}{aH}\right)^4
 \delta^2_{{\rm H}}(k)W^2(kR)\frac{dk}{k}\,.\label{eqn:sigma_defn}
\end{equation}
We will take $W(kR)$ to be a gaussian filter. The factor $10/9$ appears because 
we are interested in perturbations in the radiation era rather than the usual 
matter era.

At the end of slow-roll inflation we have $\eta = 1$, and so the power spectrum 
is
rising steeply with a spectral index $n-1 \simeq 2\eta \simeq 2$. Approximating
$\delta^2_{{\rm H}}(k)$ as a power law at the end of slow-roll inflation,
\begin{equation}
\delta^2_{{\rm H}}(k)= \delta^2_{{\rm H}} (k_{{\rm end}}) 
 \left(\frac{k}{k_{{\rm end}}}\right)^{n-1} \,, \label{eqn:powerlaw}
\end{equation}
and setting $aH = 1/R$, we can evaluate the dispersion 
Eq.~($\ref{eqn:sigma_defn}$) at horizon crossing,
\begin{equation}
\sigma^2_{{\rm hor}}(k_{{\rm end}}R) \simeq \left(\frac{10}{9}\right)^2
 \delta^2_{{\rm H}}(k_{{\rm end}})(k_{{\rm end}}R)^4 
I(n) \,,\label{eqn:sigma_eoi}
\end{equation}
where $I(n)$ is a numerical factor of order unity which depends on the spectral
index. The length scale at the end of slow-roll inflation also provides the 
natural scale
over which to smooth the power spectrum since it is the scale on which black
holes will predominantly form. Setting $k_{{\rm end}}R = 1$ we can evaluate 
$I$ to be
\begin{equation}
\label{eqn:igamma}
I = \int_{0}^{1}\tilde{k}^{(n+2)}W^2(\tilde{k}k_{{\rm end}}R)d\tilde{k} 
  = \frac{1}{2}\gamma\left[(n+3)/2,1\right] \,,
\end{equation}
where $\gamma \left[\alpha,x\right]$ is the incomplete gamma function. For 
example, with $n-1 = 3$ we have $I \simeq 0.067$, $n-1 = 2$ we have 
$I \simeq 0.080$, while for $n-1 = 1$ we have $I \simeq 0.100$.
We note that in the limit of large $n$, holding $\delta^2_{{\rm H}} (k_{{\rm 
end}})$ constant, the contribution of this spike to $\sigma_{{\rm hor}}$
is suppressed by the numerical factor $I$. We can see immediately
that under the power-law approximation
of Eq.~(\ref{eqn:powerlaw}), the dispersion $\sigma_{{\rm hor}}$
only depends weakly on the exact value of the spectral index at
the end of slow-roll inflation which we take to be a nominal
and conservative $n=3$.

The main dependence of $\sigma_{{\rm hor}}$ is on the perturbation amplitude 
$\delta_{{\rm H}}(k_{{\rm end}})$, although in our case, and for any
sharply rising power spectrum, $\delta_{{\rm H}}(k_{{\rm end}})$ depends
\emph{strongly} on the exact condition for the point where slow-roll
inflation ends. 

\begin{figure*}[t]
\centering 
\leavevmode \epsfysize=6.0cm \epsfbox{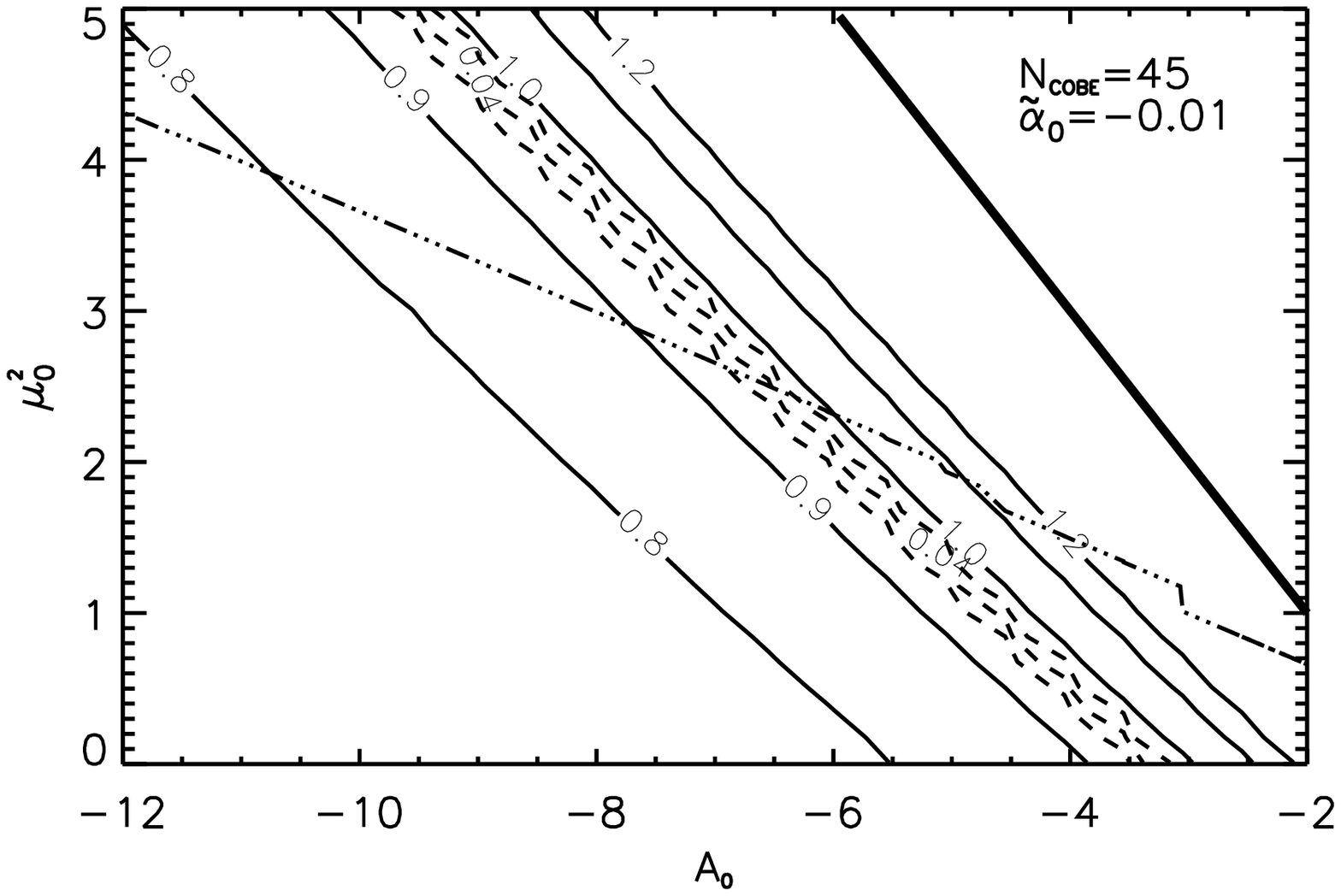}
\epsfysize=6.0cm \epsfbox{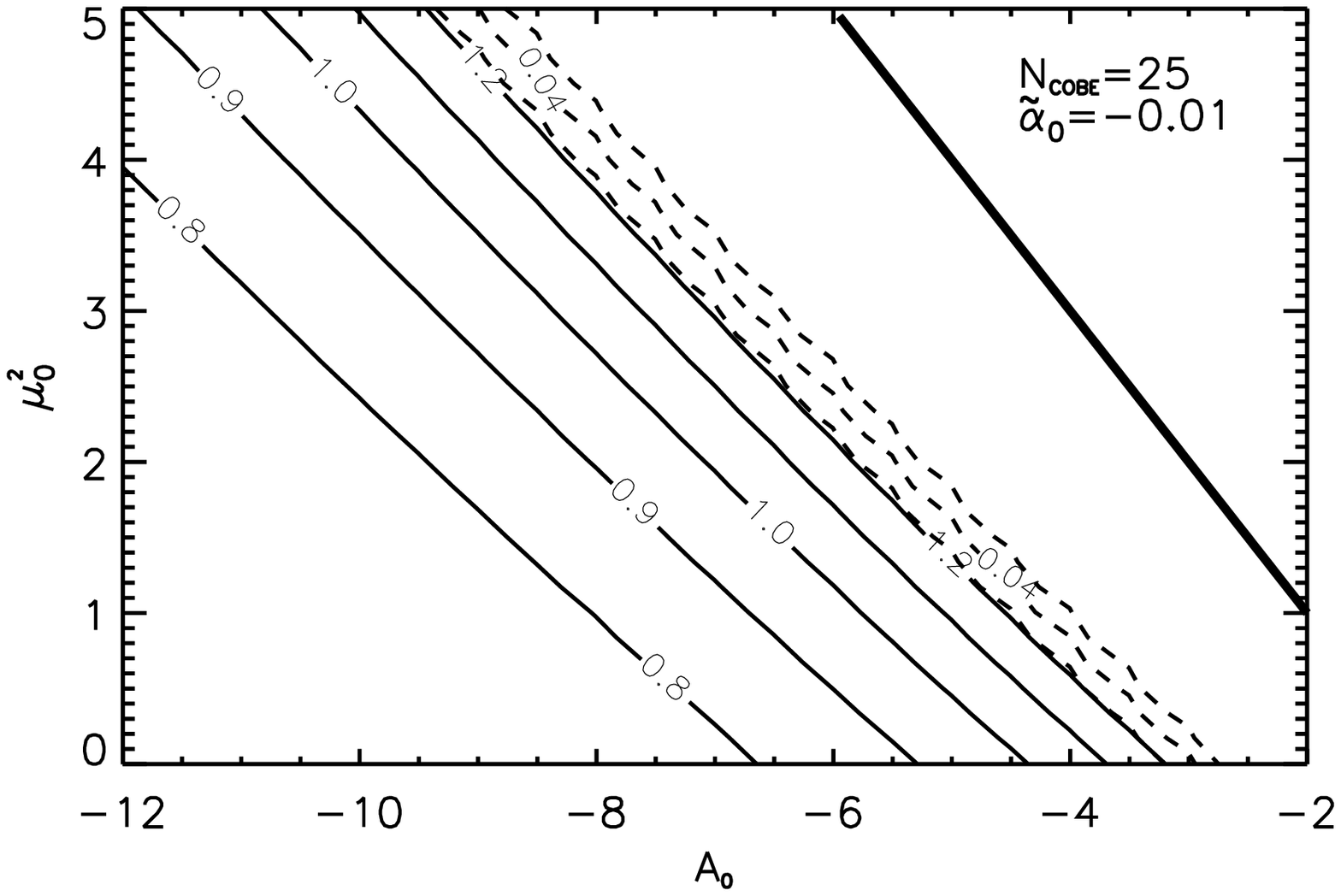} 
\caption[fig1]{\label{fig:alpha_neg} As Fig.\ref{fig:alpha_pos}, but for 
$\tilde{\alpha}_0 = -0.01$. The thick line to the
right is a bound on the allowed values of the parameters 
($\left|A_0\right| > \mu_0^2 + 1)$. The dashed lines, reading from left to 
right, are the
$\sigma_{{\rm hor}} = 0.02, 0.04, 0.08$ contours, and the region 
to the right of these contours is excluded by PBH constraints.
From Eq.~(\ref{eqn:instant_r}), instant reheating requires
$V_0/M^4_{{\rm P}} \gtrsim 10^{-36}$ for $N_{{\rm COBE}}=45$
indicated by the dot-dashed contour; the parameter space above this
contour is excluded.}
\end{figure*}

As shown in Ref.~\cite{CLLW}, the black hole constraint across all scales simply 
amounts to
\begin{equation}
\sigma_{{\rm hor}} \lesssim 0.04 \,. \label{eqn:sigma_constraint}
\end{equation}
This is sufficient to ensure that no more than $10^{-20}$ of the mass density of 
the Universe is channelled into black holes.
The constraint on $\sigma_{{\rm hor}}$ is expected to be accurate to within a
factor of 2, which is small compared to the 2--3 orders of magnitude that the
power spectrum can rise between the $\eta = 1/2$ and $\eta = 1$ 
points. This uncertainty is therefore much less important than the uncertainty 
of the end-point of inflation, which we discuss further below.

As well as the observational constraints on the model, there is a 
self-consistency constraint which must be satisfied, which is to ensure that the 
inflationary energy scale, once normalized to COBE, is high enough to permit the 
claimed number of $e$-foldings $N_{{\rm COBE}}$. If we conservatively assume 
instant reheating after inflation, and that the radiation era is not punctuated 
by episodes of thermal inflation or temporary matter domination, the upper bound 
on the number of $e$-foldings that can take place is
\begin{equation}
N_{{\rm COBE}} < 48 + \ln(V_0^{1/4}/10^{10}\mathrm{GeV}),\label{eqn:instant_r}
\end{equation}
which requires $V_0/M_{{\rm P}}^4 \gtrsim 10^{-36}, 10^{-72}$ for 
$N_{{\rm COBE}} = 45, 25$ respectively.

\subsection{Results}

The results for $N_{{\rm COBE}} = 45,25$ and $\tilde{\alpha}_0 = 0.01$
are shown in Fig.~\ref{fig:alpha_pos}.\footnote{If one
compares our contours of $n$ on COBE scales with those in Ref.~\cite{CL},
differences are apparent especially at large values of $n$. These differences 
are due to a slightly different choice for the end of inflation, and our use of 
numerical calculations rather than an approximate analytic technique. The 
differences should be regarded as indicating the arbitrariness in deciding where 
inflation comes to an end once the form Eq.~(\ref{eqn:inf_potential}) breaks 
down away from the slow-roll regime. This leads to a different identification of 
the part of the potential corresponding to COBE scales, and the running of $n$ 
causes the contours to slide to a different location. The construction of a 
complete model including a mechanism for ending inflation would be needed to 
remove this uncertainty.} We plot three
different values of $\sigma_{{\rm hor}}$; the central one is the best
guess at where the constraint lies and the others indicate the
uncertainty.  We see that the PBH constraint actually quite closely
follows lines of constant $n$ on COBE scales, enabling us to use this
to summarize the constraint.  In the case of $N_{{\rm COBE}} = 45$ we
find that a large amount of otherwise viable parameter space is
excluded: models with $n \gtrsim 1.1$ are ruled out.
This should be compared with
PBH constraints on inflation models with constant spectral index, for
which the end result is $n \geq 1.25$ are excluded~\cite{GL2}. It is
of course not surprising that the constraint on $n$ should be stronger
for the running-mass model whose spectral index increases as a
function of wavenumber $k$.

For $N_{{\rm COBE}} = 25$ the PBH constraints are less severe, models with $n 
\gtrsim 1.3$ being excluded. The simplest explanation
is that the mass runs for fewer $e$-foldings leading to a safer period of 
slow-roll inflation. In fact it is a combination of two factors that
make the $N_{{\rm COBE}} = 25$ case safer. Firstly, reducing $N_{{\rm
COBE}}$ has the effect of translating the spectral index contours away
from the region  of parameter space previously excluded: for a given
spectral index contour, the values of $A_0$ and $\mu_0^2$ must
increase as $N_{{\rm COBE}}$ decreases to ensure that the running of
the mass up to $\eta=1$ is faster. Secondly, when we renormalize the new
spectra (for given values of $\mu_0^2$ and $A_0$) to COBE, the overall 
amplitude at the end of slow-roll inflation will be reduced for all
models with $n > 1.0$ as compared to the $N_{{\rm COBE}} = 45$ case, because the 
spectral index $n$ is an increasing function of $N$. Therefore, since 
the $\sigma_{{\rm hor}} = 0.04$ contour lies in the region $n > 1.0$ 
for $N_{{\rm COBE}} = 45$, the excluded region of parameter space
shrinks for the $N_{{\rm COBE}} = 25$ case.

Next we look at the results for $\tilde{\alpha}_0 = -0.01$,
which are shown in Fig.~\ref{fig:alpha_neg}.  For $N_{{\rm COBE}} = 45$, the
PBH constraints are more severe, ruling out models with $n \gtrsim
1.0$. This result is related to the running strength throughout
inflation, given by \cite{KT} (neglecting terms in $\epsilon$)
\begin{equation}
\frac{dn}{d\ln k} \simeq -2\xi^2,\label{eqn:run_strength} 
\end{equation}
Over observable cosmological scales the running strength given by
$-\xi^2$ is generally greater for negative $\tilde{\alpha}_0$ case 
(for a given spectral index contour $n$) resulting in a larger value
of $\eta$ throughout slow-roll inflation. For instance, for $n-1=0$ we 
have, over observable scales~\cite{CLR},
\begin{equation}
\frac{dn}{d\ln k} 
\simeq8A_0^2\tilde{\alpha}_0^2\left(1+\frac{\mu_0^2}{A_0}\right)^3.
\end{equation}
Towards the end of slow-roll inflation the running becomes stronger
in the positive $\tilde{\alpha}_0$ case as the field approaches our
fixed end point, $\eta=1$. For $N_{{\rm COBE}} = 25$ the
spectral index contours are once again shifted close to the point
where the PBH constraints cease to be very interesting, with $n \gtrsim
1.3$ being excluded.

\begin{figure}[t]
\centering 
\leavevmode\epsfysize=6cm \epsfbox{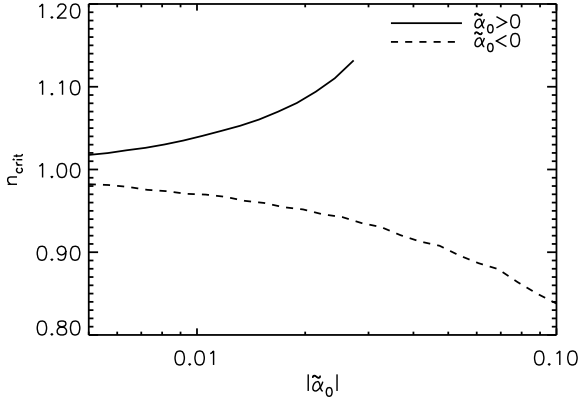}\\
\caption[fig1]{\label{fig:alpha0} The effect of varying the coupling constant,
$\tilde{\alpha}_0$ ($N_{{\rm COBE}}=45$).
$n_{{\rm crit}}$ is the spectral index on COBE scales above which the model is 
ruled out, assuming the constraint is $\sigma_{{\rm hor}} < 0.04$. For 
$\tilde{\alpha}_0>0$ the constraint is weakened as $\tilde{\alpha}_0$ is 
increased, 
while for  $\tilde{\alpha}_0<0$ the constraint becomes more restrictive as 
$\left|\tilde{\alpha}_0\right|$ is increased.}
\end{figure}

Finally we would like to know what effect varying the value of 
$\tilde{\alpha}_0$
has on the PBH constraints. Given that the $\sigma_{{\rm hor}}$ contours
are approximately parallel to the spectral index contours, $n_{{\rm COBE}}$,
we can reduce the dimensionality of this calculation, and simply assign
to each $\tilde{\alpha}_0$ some critical value of the spectral index (on COBE 
scales),
$n_{{\rm crit}}$, above which the model is excluded. We will assume the 
constraint is $\sigma_{{\rm hor}} < 0.04$. The results are shown in
Fig.~\ref{fig:alpha0} and illustrate the trends of the PBH constraints
when we vary $\tilde{\alpha}_0$. For $\tilde{\alpha}_0>0$ the PBH constraints 
become less
restrictive as $\tilde{\alpha}_0$ is increased, since the other parameters of 
the model 
take on smaller values, which has the effect of reducing the overall running 
strength. 
For $\tilde{\alpha}_0<0$ the PBH constraints become more restrictive as 
$\left|\tilde{\alpha}_0\right|$ is increased, since the other model parameters
remain fairly static, and the overall running strength becomes greater.

Using the so-called linear approximation described in Refs.~\cite{CL,LC}, the 
observational 
constraints
on the running-mass model can be expressed in terms of two new parameters
$c$ and $\sigma$, rather than directly in terms of the three model parameters
$\mu_0^2$, $A_0$ and $\tilde{\alpha}_0$. The quantity $c$ is related to the 
coupling
strengths involved and $\sigma$ is an integration constant related to the
endpoint of slow-roll inflation. In more general models these two quantities
are still enough to describe the density perturbation over cosmological 
scales, but not away from these scales where the linear approximation
breaks down, hence the need for a numerical calculation of the PBH constraints.
For given values of $N_{{\rm COBE}}$ and $n_{{\rm crit}}$, though, the 
corresponding constraints on the $c$-$\sigma$ plane can be found from 
Fig.~\ref{fig:alpha0} using the
relation~\cite{CL}
\begin{equation}
n-1=2\sigma e^{-c N_{{\rm COBE}}}-2c \,.\label{eqn:lin_approx}
\end{equation}

Before ending, we need to comment on our choice for the end of slow-roll
inflation given by Eq.~(\ref{eqn:eoi_defn}); as we have remarked the constraints
can be highly sensitive to this and we need to ensure we are being conservative.
There is no kinematical reason why inflation cannot proceed when $\epsilon \ll
1$ and $\eta \lesssim 1$, although we know that this situation can be tolerated
for no more than a few $e$-folds, given a COBE-normalized spectrum of
perturbations.  This is just restatement of the $\eta$-problem, and indeed is
observed in our simulations where inflation always proceeds to the $\eta = 1$
point and beyond.  However, as we move into the regime where $\eta \gtrsim 1$ we
find that the running strength given by $-\xi^2$ begins to blow up, marking the
failure of the one-loop approximation.  This suggests it is dangerous to try and
proceed further along the potential even though the numerical simulations show
inflation continuing and the spectrum continuing to rise. Thus, evolving the
inflaton to $\phi_{{\rm end}}$ where $\eta=1$ but not beyond appears reasonable.
Models with a larger spectral index on COBE scales will of course violate the
bound on $\sigma_{{\rm hor}}$ before $\phi_{{\rm end}}$ is reached and in this
way are more strongly constrained.

We are not able to say what happens after the end of slow-roll inflation. 
Eventually inflation is supposed to end via the hybrid mechanism when the field 
passes an instability point. However, since the form of the potential is 
breaking down by then we cannot make accurate computations in order to check 
whether there are any dangerous perturbations produced during this final era. In 
ignoring such perturbations, we are adopting a conservative approach to the 
constraints, as our constraints from the evolution up to the end of the 
slow-roll era remain valid whatever might happen subsequently.

\section{Conclusions}

We have investigated primordial black hole constraints within the context of the
failure of slow-roll inflation, focusing on the well-motivated running-mass
model of inflation which features a strong scale-dependence of the spectral 
index \cite{LC}. Although applying to the amplitude of perturbations on very 
short scales, to a
good approximation the constraint can be represented as a constraint on $n_{{\rm
COBE}}$, the spectral index on the largest observable scales.  The constraint
depends strongly on the number of $e$-foldings $N_{{\rm COBE}}$ between the
production of those perturbations and the end of inflation, and, as with models
with a constant spectral index, the constraint becomes weaker as $N_{{\rm
COBE}}$ is reduced.
 
We have shown that a significant region of the parameter space of the model, 
viable under other constraints, is excluded by excess production of black holes. 
This demonstrates the importance of evaluating the density perturbation spectrum 
not just across astrophysical scales but also right to the end of inflation. In 
models where the slow-roll approximation holds accurately only over a limited 
range of scales, such as the running-mass model, there will be strong deviations 
from scale-invariance towards short scales. In models where the deviation takes 
the form of a strongly blue spectrum, excessive black hole production is always 
likely to be a danger.

\section*{Acknowledgments}

S.M.L. and I.J.G. are supported by PPARC. We thank David Lyth for encouraging us 
to look at this problem, and Laura Covi for useful discussions. We acknowledge
the use of the Starlink computer systems at the University of Sussex and 
Imperial College. 

 

\begin{references}
\bibitem{LR} D. H. Lyth and A. Riotto, Phys. Rept. {\bf 314}, 1 (1999), 
        hep-ph/9807278. 
\bibitem{S1} E. D. Stewart, Phys. Lett. B {\bf 391}, 34 (1997),
         hep-ph/9606241.
\bibitem{S2} E. D. Stewart, Phys. Rev. D {\bf 56}, 2019 (1997),
         hep-ph/9703232.
\bibitem{CGL} E. J. Copeland, I. J. Grivell and A. R. Liddle, Mon. Not. Roy.
	Astr. Soc. {\bf 298} 1233 (1998), astro-ph/9712028.
\bibitem{CLR} L. Covi, D. H. Lyth and L. Roszkowski, Phys. Rev. D {\bf 60},
	023509 (1999), hep-ph/9809310.
\bibitem{CL} L. Covi and D. H. Lyth, Phys. Rev. D {\bf 59} (1999),
        hep-ph/9809562.
\bibitem{LC} D. H. Lyth and L. Covi, astro-ph/0002397.
\bibitem{LL} A. R. Liddle and D. H. Lyth, Phys. Rep. {\bf 231}, 1 (1993),
	 astro-ph/9303019.
\bibitem{V} A. D. Linde, Phys. Lett. B {\bf 327}, 208 (1994), astro-ph/9402031;
         A. Vilenkin, Phys. Rev. Lett. {\bf 72}, 3137 (1994), hep-th/9402085. 
\bibitem{LPB} A. R. Liddle, P. Parsons and J. D. Barrow, Phys. Rev. D {\bf 50}
	(1994), astro-ph/9408015.
\bibitem{Muk} V. F. Mukhanov, Pis'ma Zh. Eksp. Teor. Fiz. {\bf 41}, 402 (1985) 
        [Sov. Phys. JETP Lett. {\bf 41}, 493 (1985)], 
        Zh. Eksp. Teor. Fiz. {\bf 84}, 1 (1988)
        [Sov. Phys. JETP Lett. {\bf 67}, 1297 (1988)]. 
\bibitem{MFB} V. F. Mukhanov, H. A. Feldman and R. H. Brandenberger, Phys. Rep. 
	{\bf 215}, 203 (1992).
\bibitem{GL1} I. J. Grivell and A. R. Liddle, Phys. Rev. D {\bf 54}, 7191 
	(1996), astro-ph/9607096.
\bibitem{SL} E. D. Stewart and D. H. Lyth, Phys. Lett. B, {\bf 302}, 171 (1993),
	gr-qc/9302019.
\bibitem{BLW} E. F. Bunn, A. R. Liddle and M. White, Phys. Rev. D {\bf 54}, 
	5917 (1996), astro-ph/9607038.
\bibitem{GL2} B. J. Carr, Astrophys. J. {\bf 205}, 1 (1975); J. H. MacGibbon 
	and B. J. Carr, Astrophys. J. {\bf 371}, 447 (1991); A. M. Green and 
	A. R. Liddle, Phys. Rev. D {\bf 56}, 6166 (1997), astro-ph/9704251.
\bibitem{CLLW} E. J. Copeland, A. R. Liddle, J. E. Lidsey and D. Wands 
	Phys. Rev. D {\bf 58}, 063508 (1998), gr-qc/9803070.
\bibitem{KT} E. J. Copeland, E. W. Kolb, A. R. Liddle and J. E. Lidsey,
	Phys. Rev. D {\bf 49}, 1840 (1994), astro-ph/9308044; A. Kosowsky and 
	M. S. Turner, Phys. Rev. D {\bf 52}, 1739 (1995), astro-ph/9504071.
\end{references}
\end{document}